\documentclass[aoas]{imsart}

\RequirePackage{amsthm,amsmath,amsfonts,amssymb}
\RequirePackage[authoryear]{natbib}
\RequirePackage[colorlinks,citecolor=blue,urlcolor=blue]{hyperref}%% uncomment this for coloring bibliography citations and linked URLs
\RequirePackage{graphicx}%% uncomment this for including figures

\usepackage{url}
\usepackage{booktabs}
\usepackage{multirow}
\usepackage[linesnumbered,ruled,vlined]{algorithm2e}

\theoremstyle{plain}

\newtheorem{theorem}{Theorem}[section]

\startlocaldefs
\endlocaldefs

\begin{document}

\begin{frontmatter}
\title{Graph Neural Poisson Models for Supply Chain Relationship Forecasting}
% \title{A sample article title with some additional note\thanksref{T1}}
\runtitle{Graph Poisson Process for Dynamic Link Prediction}
%\thankstext{T1}{A sample of additional note to the title.}

\begin{aug}
\author[A]{\fnms{Ling}~\snm{Xiang} \ead[label=e1]{lingxiang@smail.swufe.edu.cn} },
\author[A]{\fnms{Quan}~\snm{Hu}\ead[label=e2]{quanhu@smail.swufe.edu.cn}},
\author[B]{\fnms{Xiang}~\snm{Zhang}\ead[label=e3]{xiangzhang@swufe.edu.cn}},
\author[A]{\fnms{Wei}~\snm{Lan}\ead[label=e4]{lanwei@swufe.edu.cn}}
\and
\author[A]{\fnms{Bin}~\snm{Liu} \ead[label=e5]{liubin@swufe.edu.cn}}
%%%%%%%%%%%%%%%%%%%%%%%%%%%%%%%%%%%%%%%%%%%%%%
%% Addresses                                %%
%%%%%%%%%%%%%%%%%%%%%%%%%%%%%%%%%%%%%%%%%%%%%%
\address[A]{School of Statistics and Data Science, Southwestern University of Finance and Economics \printead[presep={ ,\ }]{e1,e2,e4,e5}}

\address[B]{School of Finance,
Southwestern University of Finance and Economics\printead[presep={,\ }]{e3}}
%%%%%%%%%%%%%%%%%%%%%%%%%%%%%%%%%%%%%%%%%%%%%%%
%% Only one address is permitted per author. %%
%% Only division, organization and e-mail is %%
%% included in the address.                  %%
%% Additional information (such as           %%
%% indicating the corresponding author) can  %%
%% be included in the Acknowledgments        %%
%% section if necessary.                     %%
% \author[A]{\fnms{Ling}~\snm{Xiang}\ead[label=e1]{}},
% \author[B]{\fnms{Quan}~\snm{Hu}\ead[label=e1]{???@???}},
% \author[B]{\fnms{Xiang}~\snm{Zhang}\ead[label=e2]{???@???}},
% \author[B]{\fnms{Wei}~\snm{Lan}\ead[label=e1]{???@???}}
% \and
% \author[B]{\fnms{Bin}~\snm{Liu} \thanks{[\textbf{Corresponding author indication should be put in the Acknowledgment section if necessary.}]}\ead[label=e1]{???@???}}
% %%%%%%%%%%%%%%%%%%%%%%%%%%%%%%%%%%%%%%%%%%%%%%
% %% Addresses                                %%
% %%%%%%%%%%%%%%%%%%%%%%%%%%%%%%%%%%%%%%%%%%%%%%
% \address[A]{School of Statistics and Data Science, Southwestern University of Finance and Economics\printead[presep={,\ }]d{e1}}

% \address[B]{School of Finance,
%   Southwestern University of Finance and Economics\printead[presep={,\ }]{e2}}
\end{aug}

\begin{abstract}
In supply chain networks, firms dynamically form or dissolve partnerships to adapt to market fluctuations, posing a challenge for predicting future supply relationships. We model the occurrence of supply edges (firm $ i $ to firm $ j $) as a non-homogeneous Poisson process (NHPP), using historical event counts to estimate the Poisson intensity function up to time $ t $. However, forecasting future intensities is hindered by the limitations of historical data alone. To overcome this, we propose a novel Graph Double Exponential Smoothing (GDES) model, which integrates graph neural networks (GNNs) with a nonparametric double exponential smoothing approach to predict the probability of future supply edge formations. Recognizing the interdependent economic dynamics between upstream and downstream firms, we assume that the Poisson intensity functions of supply edges are correlated, aligning with the non-homogeneous nature of the process. Our model is interpretable, decomposing intensity increments into contributions from the current edge’s historical data and influences from neighboring edges in the supply chain network. Evaluated on a large-scale supply chain dataset with 87,969 firms, our approach achieves an AUC of 93.84\% in dynamic link prediction, demonstrating its effectiveness in capturing complex supply chain interactions for accurate forecasting. The data and code are available at \url{https://github.com/lingxiang6/dynamic_link_prediction_through_GDES.git}.
\end{abstract}


\end{frontmatter}
%%%%%%%%%%%%%%%%%%%%%%%%%%%%%%%%%%%%%%%%%%%%%%
%% Please use \tableofcontents for articles %%
%% with 50 pages and more                   %%
%%%%%%%%%%%%%%%%%%%%%%%%%%%%%%%%%%%%%%%%%%%%%%
%\tableofcontents
%%%%%%%%%%%%%%%%%%%%%%%%%%%%%%%%%%%%%%%%%%%%%%
%%%% Main text entry area:

\section{Introduction}

The supply chain can be modeled as a dynamic directed graph that it evolves over time. Procurements between upstream and downstream firms in the supply chain at time $t-1$ may or may not continue into the next time. The application of link prediction techniques to forecast these procurements has garnered increasing attention recently. Through link prediction, we can discover potential partners or predict disruptions in the supply chain, which facilitates risk management for supply chain interruptions \citep{carvalho2021supply, lu2020discovering} and enhances our understanding of macroeconomic industrial growth from a microeconomic perspective \citep{acemoglu2012network, dessertaine2022out}.

Early approaches to supply chain link prediction were relatively simplistic. For instance, \citet{wichmann2020extracting} proposed using natural language processing techniques to automatically extract supply chain information from the Internet to infer potential supplier relationships, relying solely on node features for link prediction. Similarly, \citet{mungo2023reconstructing} collected financial and geographical features to predict links between firms. However, \citet{brintrup2018research} argued that link prediction should focus more on network topology rather than just node attributes and developped a Supply Network Link Predictor (SNLP) method to infer supplier interdependencies in supply networks using topological data and features like supplier links and product outsourcing associations. 
Recently, many studies have begun encoding domain-specific knowledge about supply chains using tools such as Graph Neural Networks (GNNs) and Knowledge Graphs to address supply link prediction problems \citep{kosasih2022machine, brockmann2022supply,liu2023interpret, kosasih2024towards}.
However, most existing work focuses on a single snapshot of the network, with little consideration given to dynamic network scenarios \citep{kosasih2022machine, brockmann2022supply, kosasih2024towards,tan2025modeling}. 

To model a supply chain connecting firms $i$ and $j$, we use a random process $A(\tau)$ to describe the occurrence of a supply event: $ A_{ij}(\tau) = 1 $ indicates that firm $ i $ provides products or services to firm $ j $ at time $\tau$, while $ A_{ij}(\tau) = 0 $ indicates a temporary disconnection. Thus, $N_{ij}(t) = \sum_{\tau=1}^{t} A_{ij}(\tau) $ represents the number of times the supply chain between $ i $ and $ j $ occurs, forming a counting process. 
We model the historical records $N(t)$ of each supply edge as a non - homogeneous Poisson process, where $N(t)$ is governed by a time-varying intensity $\lambda_{ij}(t)$. Given the upstream-downstream supply relationships, we assume that the counting processes $N_{ij}(t)$ of all supply chains are interdependent. This interdependence implies that their intensities $\lambda_{ij}(t)$ mutually affect one another. 
Therefore, we employ GNNs to capture the correlation among interconnected supply chains.

Specifically, we designed a graph double exponential smoothing algorithm (GDES) to estimate the intensity function $\lambda_{ij}(t)$ for each supply edge. At time $\tau$, the double exponential smoothed estimate of the intensity function $\lambda_{ij}(t)$ comprises three components: 1) the current one-step exponential smoothing estimate, 2) the previous two-step double exponential smoothing estimate, and 3) the double exponential smoothed estimates from neighboring supply chains. The neighboring estimations are characterized by GNNs, integrating textual features of the firms and transportation costs between them.
We found that this extended double exponential smoothing based on the supply chain graph is a generalization of traditional exponential smoothing techniques, where multivariate double exponential smoothing is a special case.

The contributions of this paper can be summarized as follows: 1) We constructed a graph-based non-homogeneous Poisson process model for dynamic prediction of supply chain events.
2) We extended the double exponential smoothing estimation within a graph framework.
3) Our experiments on real-world datasets demonstrated that the accuracy of predicting future supply chain disruptions and occurrences AUC exceeded 93\%.

The remainder of the article is organized as follows: Section 2 presents the preliminaries about graph neural networks. In Section 3 a graph double exponential smoothing method was proposed for supply relationship prediction. Section 4
 studies the performance of our model. Finally, Section 5 concludes.

\section{Graph neural networks: from undirected to directed graphs}

\subsection{General formulation of GNNs on undirected graphs}

Graph Neural Networks (GNNs) are designed to learn node representations by fusing graph topological structures and node attributes. For undirected graphs, the core mechanism relies on symmetric message passing, which can be formally defined using the node set and adjacency matrix.

Formally, an undirected graph is represented as \( G = (\mathcal{V}, \mathbf{A}), \mathbf{X} \), where:
\( \mathcal{V} = \{v_1, v_2, \ldots, v_n\} \) is the set of \( n \) nodes;
\( \mathbf{A} \in \mathbb{R}^{n \times n} \) is the symmetric adjacency matrix, where \( \mathbf{A}_{ij} = \mathbf{A}_{ji} = 1 \) if there is an edge between \( v_i \) and \( v_j \), and \( \mathbf{A}_{ij} = \mathbf{A}_{ji} = 0 \) otherwise (self-loops may be included by setting \( \mathbf{A}_{ii} = 1 \)).

Each node \( v_i \) is associated with an initial feature vector \( \mathbf{x}_i \in \mathbb{R}^F \). GNNs update node representations through stacked layers, where the representation of node \( v_i \) at layer \( l \) (denoted as \( \mathbf{h}_i^{(l)} \)) is computed by aggregating information from its neighbors—identified by non-zero entries in the \( i \)-th row of \( \mathbf{A} \):

\begin{equation}\label{eq:gnn}
    \mathbf{h}_i^{(l)} = \sigma\left(  \mathbf{h}_i^{(l-1)}, \text{AGG}^{(l)}\left( \{\mathbf{h}_j^{(l-1)} \mid j\in \mathcal{N}(i)\} \right) ;\mathbf{W}^{(l)} \right)
\end{equation}
where $\mathcal{N}(i):=\{j| \mathbf{A}_{ij}=1\}$ is the neighbor set of node $i$, $\mathbf{h}_i^{(0)}=\mathbf{x}_i$, \( \text{AGG}^{(l)} \) denotes the aggregation function (e.g., mean, sum, or attention-based) that combines representations of neighbors identified by \( \mathbf{A} \); \( \mathbf{W}^{(l)} \) are learnable parameters for layer \( l \); \( \sigma(\cdot) \) is a non-linear activation function (e.g., ReLU).

The matrix form of Eq.~(\ref{eq:gnn}) is,
\begin{equation}
    \mathbf{H}^{(l)}=\text{GNN}_l(\mathbf{H}^{(l-1)}, \mathbf{A};\mathbf{W}^{(l)}),
\end{equation}
where $\mathbf{H}^{(l)}=[\mathbf{h}_1^{(l)},...,\mathbf{h}_n^{(l)}]$.

Notable examples include Graph Convolutional Networks (GCNs) \cite{kipf2016semi}, which use a normalized version of \( \mathbf{A} \) for localized convolutions, and GraphSAGE \cite{hamilton2017inductive}, which leverages \( \mathbf{A} \) to sample neighbors for inductive learning. These models inherently rely on the symmetry of \( \mathbf{A} \), making them suitable for undirected graphs.

\subsection{Directed GNNs}

Directed graphs are represented as \( G = (\mathcal{V}, \mathbf{A}) \) where \( \mathbf{A} \in \mathbb{R}^{n \times n} \) is asymmetric: \( \mathbf{A}_{ij} = 1 \) indicates a directed edge from node \(i \) to node \( j \), but \( \mathbf{A}_{ji} \neq \mathbf{A}_{ij} \) in general. This asymmetry (distinguishing in-edges \( \{\mathbf{A}_{ji} = 1\} \) and out-edges \( \{\mathbf{A}_{ij} = 1\} \)) poses challenges for traditional GNNs, which assume symmetric \( \mathbf{A} \).

To address this, \cite{tan2022asymmetric} proposed Asymmetric Self-Supervised Graph Neural Networks (Asymm-SSGNN), a directed graph-specific variant with two key designs:

Asymmetric Message Passing with Directed Adjacency: The model explicitly distinguishes in-neighbors (nodes \( v_j \) with \( \mathbf{A}_{ji} = 1 \)) and out-neighbors (nodes \( v_j \) with \( \mathbf{A}_{ij} = 1 \)) using \( \mathbf{A} \). For each node \( v_i \), separate in-representation \( \mathbf{h}_i^{\text{in}} \) and out-representation \( \mathbf{h}_i^{\text{out}} \) are computed:
   \begin{align}
   \mathbf{h}_i^{\text{in}} &= \sigma\left( \mathbf{W}_{\text{in}} \cdot \text{AGG}_{\text{in}}\left( \{\mathbf{h}_j^{\text{out}} \mid \mathbf{A}_{ji} = 1\} \right) \right) \\
   \mathbf{h}_i^{\text{out}} &= \sigma\left( \mathbf{W}_{\text{out}} \cdot \text{AGG}_{\text{out}}\left( \{\mathbf{h}_j^{\text{in}} \mid \mathbf{A}_{ij} = 1\} \right) \right)
   \end{align}
   Here, \( \text{AGG}_{\text{in}} \) and \( \text{AGG}_{\text{out}} \) are asymmetric aggregators with distinct parameters, capturing direction-specific patterns encoded in \( \mathbf{A} \).

\section{Framework and model}
The notation used in this paper is summarized in Table \ref{tab:notations}.
\begin{table*}[]
\centering
\resizebox{1.0\textwidth}{!}{%
\begin{tabular}{@{}lll@{}}
\toprule
notation & variable description & \\ \midrule
$\tau$    &  time index                    &    \\
$i,j$    &    matrix indexes                  &    \\
$t$     &  current time                    &    \\
% $a_{ij}({\tau})$     &   an observation of $A_{ij}({\tau})$                 &    \\
$A_{ij}({\tau})$     &   random variable: firm $i$ provides services or sells products to firm $j $                  &    \\
$\mathbf{A}({\tau})$         & dynamic adjacent matrix at time $\tau$, its elements are observation of $A_{ij}({\tau})$                   &    \\
$N(t)$         &          Poisson counting            &    \\
$n$     & number of firms (nodes)                     &    \\
$\mathbf{S}$         & supplying embeddings of firms                   &    \\
$\mathbf{R}$         & demanding embeddings of firms                    &    \\
    $\widetilde{\mathbf{T}}({t}), \widetilde{\mathbf{T}}({t})$     &                 trend term     &    \\
    $\widetilde{\mathbf{L}}({t}),\mathbf{L}({t})$     &        level term              &    \\
$\mathbf{\mathcal{X}}$     &      firm text feature  &    \\
$\lambda_{ij}({t})$      &     intensity function                 &    \\
$\boldsymbol{\lambda}$       & intensity vector                   &    \\
$\widetilde{\boldsymbol{\lambda}}^{(1)}$, $\widetilde{\boldsymbol{\lambda}}^{(2)}$      & intermediate values of double exponential smoothing          &    \\
$\hat{\boldsymbol{\lambda}}$      &    estimates of the non-homogeneous Poisson process                  &    \\
$\breve{\boldsymbol{\lambda}}$      &   graph double exponential smoothing estimates                  &    \\
$\mathbf{X}$     &     feature matrix                 &    \\
\bottomrule
\end{tabular}%
 }
\caption{Notation list.}
\label{tab:notations}
\end{table*}

\subsection{Multivariate double exponential smoothing}
% \subsection{Multivariate Double Exponential Smoothing}
\label{sec:mdes}
% Now we review the steps of multivariate double exponential smoothing (MDES), laying the foundation for the upcoming multivariate interactive exponential smoothing. 
Multivariate Double Exponential Smoothing (MDES), as introduced by \citep{PFEFFERMANN198983}, extends the conventional univariate double exponential smoothing framework to accommodate multivariate time series data. In its univariate formulation, double exponential smoothing, commonly referred to as Holt's linear trend method, augments simple exponential smoothing by incorporating both a level and a trend component. For a univariate time series $\{y(\tau), \tau = 1, 2, \dots, t\}$, this method employs two smoothing parameters, $L(\tau)$ to govern the level and $T(\tau)$ to regulate the trend, as described below:
\begin{equation}
y(\tau + m) = L(\tau) + m \cdot T(\tau) + \epsilon(\tau),
\label{eq:univariate_r}
\end{equation}
where $m$ is a horizon, and $\epsilon(\tau)$ is the irregular component assumed to be a random disturbance with zero mean and constant variance. 
The smoothing process is as follows:
\begin{equation}
    {s}^{(1)}(\tau) = \alpha y_{\tau} + (1 - \alpha) {s}^{(1)}(\tau-1), \quad
    {s}^{(2)}(\tau) = \alpha s^{(1)}(\tau) + (1 - \alpha) {s}^{(2)}(\tau-1),
    \label{eq:univariate_p}
\end{equation}
where ${s}^{(1)}(\tau-1)$ and ${s}^{(2)}(\tau-1)$ are the smoothed values at time $\tau-1$.
Then the forecast for future values at horizon $m$ can be obtained (Appendix A.1):
\begin{equation}
    \hat{y}(\tau + m) = L(\tau) + m \cdot T(\tau).
    \label{eq:univariate_estimate}
\end{equation}

MDES can be derived based on Eq.~(\ref{eq:univariate_r}) -(\ref{eq:univariate_estimate}) as follows, 
% \begin{equation*}
% \mathbf{y}(\tau + m) = \mathbf{L}(\tau) + m \cdot \mathbf{T}(\tau) + \mathbf{\epsilon}(\tau),
% \label{eq:MDES_r}
% \end{equation*}
\begin{equation}
    \begin{aligned}
    \mathbf{y}(\tau + m) &= \mathbf{L}(\tau) + m \cdot \mathbf{T}(\tau) + \mathbf{\epsilon}(\tau),\\
    \mathbf{s}^{(1)}(\tau) &= \alpha \mathbf{y}_{\tau} + (1 - \alpha) \mathbf{s}^{(1)}(\tau-1), \\
    \mathbf{s}^{(2)}(\tau) &= \alpha \mathbf{s}_{\tau}^{(1)} + (1 - \alpha) \mathbf{s}^{(2)}(\tau-1).
    \end{aligned}
    \label{eq:MDES_p}
\end{equation}
Then the forecast of the future data at horizon $m$ is,
\begin{equation}
    \hat{\mathbf{y}}(\tau+m) = \mathbf{L}(\tau) + m \cdot \mathbf{T}(\tau).
    \label{eq:MDES_estimate}
\end{equation}
where $\mathbf{L}(\tau) = 2\mathbf{s}^{(1)}(\tau) - \mathbf{s}^{(2)}(\tau)$, and $\mathbf{T}(\tau)  = \frac{\alpha}{1-\alpha}(\mathbf{s}^{(1)}(\tau) - \mathbf{s}^{(2)}(\tau))$. 
Further details are provided in Appendix A.2.

\subsection{Formation}A supply chain network can be conceptualized as a directed dynamic graph $G(V, \mathbf{A}(\tau), \mathbf{X})$, where $V$ represents a set of $n$ firms, $\mathbf{A}(\tau) \in \{0,1\}^{n \times n}$ encapsulates the dynamic procurement relationships among them~\citep{choi2001supply}, and $\mathbf{X} \in \mathbb{R}^{n \times d}$ denotes the feature matrix of firms. Specifically, the $(i,j)$-th element of $\mathbf{A}(\tau)$ is exactly the random process $A_{ij}(\tau)$ defined before, equals 1 if firm $j$ procures goods or services from firm $i$ at time $\tau$, and 0 otherwise.

\begin{figure*}[!thp]
    \centering
    \includegraphics[scale=1.1]{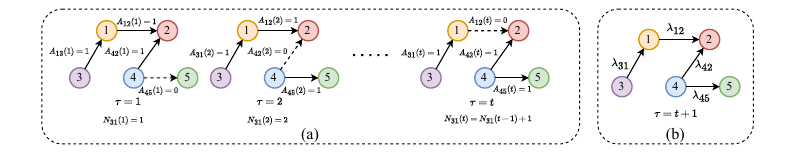} 
    \caption{A dynamic supply chain. (a): illustrates that occurrences on the supply side over time can be modeled as a counting process. (b): modeling this counting process using the intensity function of a non-homogeneous Poisson process.}
    \label{fig:nhpe_process}
\end{figure*}

\subsection{Non-homogeneous Poisson estimation (NHPE)}
\subsubsection{Counting process and its Intensity $\lambda$}
% Procurement events within the supply chain, from firm $ i $ to firm $ j $, can be modeled as a stochastic process, characterized by the random variable $ A_{ij}(\tau) $. 
$ A_{ij}(\tau) $ represents a single observation at time $ \tau $, where $ \tau = 1, 2, \ldots, t $ and $ i, j \in \{1, 2, \ldots, n\} $. We formly define the counting process $N_{ij}(t)$ based on $ A_{ij}(\tau) $ at the current time $ t $ as follows:
\begin{equation*}
    N_{ij}(t) = A_{ij}(1)+A_{ij}(2)+\cdots + A_{ij}(t).
\end{equation*}
Figure~\ref{fig:nhpe_process} (a) illustrates a demo of $ N_{ij}(t)$. For instance, at time $ \tau = 1 $, $ A_{12}(1) = 1 $ indicates that firm 1 supplies products or services to firm 2. Similarly, $ A_{42}(1) = 1 $ denotes that firm 4 provides products or services to firm 2, while $ A_{45}(1) = 0 $ signifies no procurement between firms 4 and 5 at $ \tau = 1 $. At time $ \tau = 2 $, we observe $ A_{12}(2) = 1 $ and $ A_{45}(2) = 1 $. Consequently, the counting process for edges, such as (1,2) and (4,5), yields $ N_{12}(2) = A_{12}(1) + A_{12}(2) = 2 $ and $ N_{45}(2) = A_{45}(1) + A_{45}(2) = 1 $.

In this study, we model the counting process $ N_{ij}(t) $ for a supply chain edge $ i \rightarrow j $ as a Poisson process. Its intensity function, denoted $ \lambda_{ij}(t \mid H_t) := \lambda_{i \rightarrow j}(t \mid H_t) $, represents the conditional rate of an event occurring at time $ t $, given the event history $ H_t := \{N_{ij}(1), N_{ij}(2), \ldots, N_{ij}(t)\} $, which encompasses all events prior to time $ t $ ~\citep{omi2019fully}. The intensity function is formally defined as:
\begin{equation*}
    \lambda_{ij}(t \mid H_t) = \lim_{\Delta t \to 0} \frac{P(A_{ij}(\tau) = 1, \tau \in [t, t + \Delta t] \mid H_t)}{\Delta t}.
\end{equation*}
In subsequent discussions, the event history $ H_t $ will be omitted for brevity, such that $ \lambda_{ij}(t) := \lambda_{ij}(t \mid H_t) $.
\begin{figure*}[!htp]
    \centering
    \includegraphics[scale=0.8]{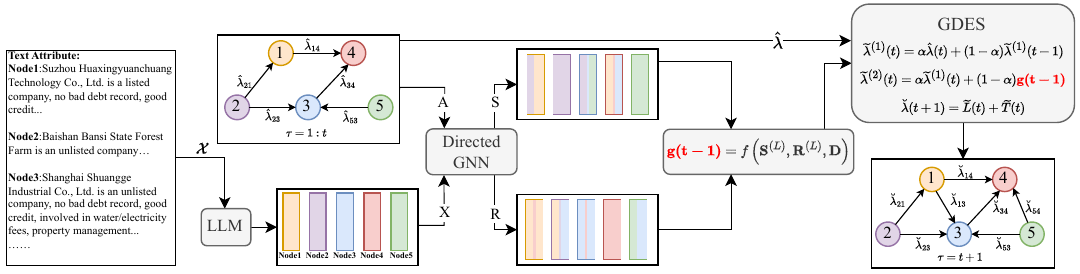} 
    \caption{The architecture of GDES. 
    % we first use the historical data to obtain    $\hat{\boldsymbol{\lambda}}(16)$. In the calculation of $\mathbf{g}(15)$, we embed the text using the Qwen and BERT to estimate $\hat{\boldsymbol{\lambda}}(16)$, then obtaining the estimated value 
    % $\breve{\boldsymbol{\lambda}}(17)$.
    }
    \label{fig:wide-figure}
\end{figure*}

\subsubsection{Fitting $\lambda$}
\label{sec:estimateLambda(t)}
Given the intricate economic dynamics of supply chains, we model the counting process $ N_{ij}(t) $ as a non-homogeneous Poisson process with a time-varying intensity function $ \lambda_{ij}(t) $. As shown in Eq. (\ref{eq:NHPE_estimate}), the estimated intensity function $ \hat{\lambda}_{ij}(t) $ is derived from the event history $ H_t $. This estimation leverages the Poisson probability function to compute $ \hat{\lambda}_{ij}(t) $ based on the observed event count $ N_{ij}(t) $,
\begin{equation}
    \hat{\lambda}_{ij}({t}) \leftarrow \text{Poisson}\left ( \frac{(\lambda_{ij}({t}))^{N_{ij}(t)}}{N_{ij}(t)!}e^{-\lambda_{ij}({t})} \Big| N_{ij}(t)\right ).
    \label{eq:NHPE_estimate}
\end{equation}
In practice, various methods exist for estimating $\hat{\lambda}_{ij}(t)$. In this study, we adopt linear functions, sigmoid functions, and Gaussian kernel functions \citep{goulding2016event}  for the estimation, as outlined below:
\begin{enumerate}
    \item Linear expansion method. \begin{equation}
    \hat{\lambda}_{ij}(t) = \sum_{k=1}^{t-1}a_{k} \cdot N_{ij}(k) + b_{k},
    \label{eq:linear}
\end{equation}
    \item Sigmoid expansion method.
    \begin{equation}
    \hat{\lambda}_{ij}(t) = \sum_{k=1}^{t-1}a_{k} \cdot sigmoid(N_{ij}(k)), 
    \label{eq:sigmoid}
\end{equation}
    \item Gaussian expansion method.
    \begin{equation}
    \hat{\lambda}_{ij}(t) = \sum_{k=1}^{t-1}a_{k} \cdot B_{k}(N_{ij}(k)), 
    \label{eq:gaussian}
\end{equation}
\end{enumerate}
where $a_k$ are the parameters, 
\begin{equation*}
    B_{k}(N_{ij}(k)) = \frac{1}{\sigma \cdot \sqrt{2\pi}}e^{-\frac{(N_{ij}(k)-u_{k})^{2}}{2\sigma^{2}}}
\end{equation*}
is a Gaussian kernel function, $u_{k} = \frac{(k-1)\eta}{(t-1)-1}, \sigma = \frac{\eta}{k}$,
and $\eta = \frac{3}{2}$ is used in our method to fit $\lambda$ up to the current time $t$.

With the event history $ H(t) := \{N(1), N(2), \ldots, N(t)\} $, we can estimate the intensity functions $ \{\hat{\lambda}(\tau)\}_{\tau=1}^t $ using the three non-homogeneous Poisson process methods described above. Our goal is to predict the intensity $ \hat{\lambda}(t+1) $ for link prediction. However, estimating $ \hat{\lambda}(\tau) $ for earlier times $ \tau $ may lead to a cold-start problem due to limited data. Additionally, the dynamic estimation process is computationally complex, and the resulting experimental outcomes are intricate. To address these challenges, we focus our estimation solely on the intensity function at the current time $ t $, i.e., fitting $ \hat{\lambda}(t) $ using $ H(t) $.
The fitting process for $ \hat{\lambda}_{ij}(t) $ using the event history $ H(t) $, as specified in Eqs. (\ref{eq:linear})–(\ref{eq:gaussian}), can be optimized by maximizing the Poisson log-likelihood function,
\begin{equation*}
    -\sum_{\forall i, j} \left( N_{ij}(t) \log(\hat{\lambda}_{ij}(t)) - \hat{\lambda}_{ij}(t) - \log(N_{ij}(t)!) \right).
    \label{eq:loss1_2}
\end{equation*}

\subsection{Graph double exponential smoothing}
We can directly estimate $ \hat{\lambda}(t) $ using the event history $H(t)$, as detailed in Section~\ref{sec:estimateLambda(t)}\footnote{For clarity, subscripts $ i $ and $ j $ are omitted in this context.}. However, at time $ t+1 $, we aim to estimate $ \lambda(t+1) $ for link prediction. Estimating $ \hat{\lambda}(t+1) $ using the same approach as $ \hat{\lambda}(t) $ is infeasible due to the absence of $ N(t+1) $, the target variable for prediction.

To simultaneously analyze all edges within the supply chain network, we define a vector $\boldsymbol{\lambda} $, where each component $\lambda_{i \rightarrow j} $ represents a directed edge from node $i $ to node $j $. The dimensionality of $\boldsymbol{\lambda} $ corresponds to the total number of directed edges in the supply chain network $G $.

Building on this, we propose a novel methodology, termed Graph Double Exponential Smoothing (GDES), which extends the double exponential smoothing framework outlined in Section~\ref{sec:mdes}. This approach facilitates the estimation of $\boldsymbol{\lambda}(t+1) $ as follows:
\begin{equation*}
    \breve{\boldsymbol{\lambda}}(t+1) = \widetilde{\mathbf{L}}(t) + \widetilde{\mathbf{T}}(t) + \widetilde{\mathbf{\epsilon}}(t),
    \label{eq:GDES_r}
\end{equation*}
where the level term $ \widetilde{\mathbf{L}}(t) $ and trend term $ \widetilde{\mathbf{T}}(t) $ are defined as:
% \begin{equation}\label{eq:L(t) and T(t)}
% \widetilde{\mathbf{L}}(t) = 2 \widetilde{\boldsymbol{\lambda}}^{(1)}(t) - \widetilde{\boldsymbol{\lambda}}^{(2)}(t),   \\    \widetilde{\mathbf{T}}(t) = \frac{\alpha}{1 - \alpha} \left( \widetilde{\boldsymbol{\lambda}}^{(1)}(t) - \widetilde{\boldsymbol{\lambda}}^{(2)}(t) \right),
% \end{equation}
\begin{equation}
    \begin{aligned}
        \widetilde{\mathbf{L}}(t) &= 2 \widetilde{\boldsymbol{\lambda}}^{(1)}(t) - \widetilde{\boldsymbol{\lambda}}^{(2)}(t), \\
        \widetilde{\mathbf{T}}(t) &= \frac{\alpha}{1 - \alpha} \left( \widetilde{\boldsymbol{\lambda}}^{(1)}(t) - \widetilde{\boldsymbol{\lambda}}^{(2)}(t) \right),
    \end{aligned}
    \label{eq:L(t) and T(t)}
\end{equation}
and the one-step and two-step smoothed terms, $ \widetilde{\boldsymbol{\lambda}}^{(1)}(t) $ and $ \widetilde{\boldsymbol{\lambda}}^{(2)}(t) $, are given by:
% \begin{equation} \label{eq:GDES_p}
%         \widetilde{\boldsymbol{\lambda}}^{(1)}(t) = \alpha \hat{\boldsymbol{\lambda}}(t) + (1 - \alpha) \widetilde{\boldsymbol{\lambda}}^{(1)}(t-1), \\
%         \widetilde{\boldsymbol{\lambda}}^{(2)}(t) = \alpha \widetilde{\boldsymbol{\lambda}}^{(1)}(t) + (1 - \alpha) \mathbf{g}(t-1).
% \end{equation}
\begin{equation}
    \begin{aligned}
        \widetilde{\boldsymbol{\lambda}}^{(1)}(t) &= \alpha \hat{\boldsymbol{\lambda}}(t) + (1 - \alpha) \widetilde{\boldsymbol{\lambda}}^{(1)}(t-1), \\
        \widetilde{\boldsymbol{\lambda}}^{(2)}(t) &= \alpha \widetilde{\boldsymbol{\lambda}}^{(1)}(t) + (1 - \alpha) \mathbf{g}(t-1).
    \end{aligned}
    \label{eq:GDES_p}
\end{equation}
Here, $ \mathbf{g}(t-1) $ represents a \textbf{GDES term} that captures two key factors: 1) \textit{the intrinsic intensity increment of a firm} and 2) \textit{the influence of neighboring firms within the supply chain network on the current node's intensity increment}. To model these factors, we employ a GNN with inputs comprising the supply chain network structure and textual features of the firms. From an economic perspective, these inputs are critical for accurately computing intensity function increments.

\begin{algorithm}[H]
\label{alg:gdes}
\SetAlgoLined
\KwIn{$H(t)$}
\KwData{Initialize $\widetilde{\boldsymbol{\lambda}}^{(1)}(t-1)$}
$\hat{\boldsymbol{\lambda}}(t) \leftarrow \text{NHPE}(N(1), \dots, N(t-1), N(t))$\;
$\widetilde{\boldsymbol{\lambda}}^{(1)}(t) \leftarrow \alpha \cdot \hat{\boldsymbol{\lambda}}(t) + (1 - \alpha) \cdot \widetilde{\boldsymbol{\lambda}}^{(1)}(t-1)$\;
$\widetilde{\boldsymbol{\lambda}}^{(2)}(t) \leftarrow \alpha \cdot \widetilde{\boldsymbol{\lambda}}^{(1)}(t) + (1 - \alpha) \cdot \mathbf{g}(t-1)$\;
$\breve{\boldsymbol{\lambda}}(t+1) \leftarrow \widetilde{\mathbf{L}}(t) + \widetilde{\mathbf{T}}(t)$\;
$\mathcal{L}_{CE} \leftarrow CE(\breve{\boldsymbol{\lambda}}(t+1), \mathbf{A}(t+1))$\;
\KwRet{$\mathcal{L} = \mathcal{L}_{CE} + \gamma \cdot \|\mathbf{W}\|_2^2$}\;
\caption{Overall procedure}
\end{algorithm}  

\subsection{Loss function}
% In the non-homogeneous Poisson estimation process, we use mean squared error(MSE) as the loss function to measure the difference between the estimated value $\hat{\boldsymbol{\lambda}}(t)$ and the counted value 
% $\mathbf{N}(t)$, aiming to optimize the accuracy of the estimation, 

For the proposed GDES method, $\breve{\boldsymbol{\lambda}}_{ij}(t+1)$ is exactly the probability of the occurrence of an event $A_{ij}(\tau)$ at time $t+1$, therefore we employ the cross-entropy loss function for model optimization,
% \begin{equation}
%     \begin{aligned}
%         L &= \prod \mathbf{P}(\breve{\boldsymbol{\lambda}}(t+1), \mathbf{A}(t+1)) 
%          \\&= \prod_{\forall i \rightarrow j}(\breve{\lambda}_{ij}(t+1))^{{a}_{ij}(t+1)} \\&
%          \cdot (\breve{\lambda}_{ij}(t+1))^{1 - {a}_{ij}(t+1)}
%     \end{aligned}
%     \label{eq:P}
% \end{equation}
% \begin{equation}\label{eq:loss_MLE}
%         \mathcal{L}_{MLE} = - \sum_{\forall i \rightarrow j}{{A}_{ij}(t+1)} \cdot log(\breve{\boldsymbol{\lambda}}_{ij}(t+1)) - ({1 - {A}_{ij}(t+1)}) \cdt log(\breve{\boldsymbol{\lambda}}_{ij}(t+1)).
% \end{equation}
\begin{equation}\label{eq:loss2}
    \mathcal{L}_{CE} =- \sum_{ i \rightarrow j}CE\left({A}_{ij}(t+1), \breve{\boldsymbol{\lambda}}_{ij}(t+1) \right).
\end{equation}
To mitigate the risk of overfitting, we incorporate an $\mathcal{L}_{2}$ regularization term into the model, thereby improving its generalization capabilities.
% \begin{equation}
%     \mathcal{L}_{2} = \|\mathbf{W}\|_2^2.
%     \label{eq:loss3}
% \end{equation}
The overall loss function is
% \begin{equation}
%     \mathcal{L}_{all} = \mathcal{L}_{CE} + \mathcal{L}_{2},
%     \label{eq:total_loss1}
% \end{equation}
\begin{equation}
    \mathcal{L} = \mathcal{L}_{CE} + \gamma \cdot \|\mathbf{W}\|_2^2,
    \label{eq:total_loss}
\end{equation}
as stated in Theorem \ref{thm:Supply}, $\mathbf{W}$ represents the parameter matrices of GNNs.
% \textcolor{red}{[we have to clarify what the $\mathbf{W}$ refers to, we have too many parameters in our model.]}
The entire process is illustrated in Figure~\ref{fig:wide-figure}, while the computational procedures are detailed in Algorithm~\ref{alg:gdes} and Algorithm~\ref{alg:processing_algorithm}. For our analysis, we set the parameter to $t = 16$.

Algorithm~\ref{alg:gdes} summarizes the GDES methodology. Our approach is based on the premise that the increment of $ \lambda(t+1) $ for each supply edge is influenced not only by its own historical data but also by interactions among edges within the network. By leveraging GNNs to capture these complex interaction patterns, we enhance the estimation of $ \lambda(t+1) $, yielding the corresponding estimate $ \breve{\boldsymbol{\lambda}}(t+1) $\footnote{Note that $ \breve{\lambda} $ and $ \hat{\lambda} $ are derived using distinct estimation approaches and are therefore different.}.

\subsection{GDES term $ \mathbf{g}(t-1) $}
This section elaborates on the computation of the GDES term $ \mathbf{g}(t-1) $. As previously discussed, $ \mathbf{g}(t-1) $ encapsulates the incremental information of the intensity function, comprising two primary components: (1) the intensity increment of the current supply edge and (2) the influence of neighboring edges within the supply chain network.

To model the interactions between the current edge and its neighboring edges, we employ a Directed Graph Neural Network (DGNN) model \citep{tan2022asymmetric}. Specifically, we define the supplying embedding $ \mathbf{S} \in \mathbb{R}^{n \times p} $ and the demanding embedding $ \mathbf{R} \in \mathbb{R}^{n \times p} $ for each firm in the supply chain. These embeddings are learned using the DGNN, with initial conditions set as $ \mathbf{S}^{(0)}, \mathbf{R}^{(0)} \leftarrow \mathbf{X} $, and updated as follows:
\begin{equation}
    \begin{aligned}
        \mathbf{S}^{(l)} = \text{DGNN}_l \left( \mathbf{S}^{(l-1)}, \mathbf{A}_{\text{out}}(t-1) \right), \\
        \mathbf{R}^{(l)} = \text{DGNN}_l \left( \mathbf{R}^{(l-1)}, \mathbf{A}_{\text{in}}(t-1) \right),
    \end{aligned}
    \label{eq:feature_update}
\end{equation}
where $ l = 1, \ldots, L $ denotes the DGNN layer, and $ \mathbf{A}_{\text{out}}(t-1) $ and $ \mathbf{A}_{\text{in}}(t-1) $ are the adjacency matrices of direct predecessors and successors, respectively, derived from $ \mathbf{A}(t-1) $ \citep{tan2022asymmetric}. Subsequently, we compute a match matrix $ \mathbf{M} $ as:
\begin{equation*}
    \mathbf{M}_{ij} = \mathbf{S}^{(L)}_i (\mathbf{R}^{(L)}_j)^{\top},
\end{equation*}
where $ \mathbf{M}_{ij} $ quantifies the compatibility between the supply characteristics of firm $ i $ and the demand characteristics of firm $ j $.

While $ \mathbf{M} $ captures the inherent alignment between the supply and demand attributes of firms $ i $ and $ j $, practical supply chain decisions, such as selecting upstream or downstream partners, often require consideration of external factors, notably transportation costs. To address this, we formulate the GDES term $ \mathbf{g}(t-1) $ as a function of three factors: the supply embedding $ \mathbf{S} $, the demand embedding $ \mathbf{R} $, and the transportation cost matrix $ \mathbf{D} $. Specifically, $ \mathbf{g}(t-1) $ integrates both the demand compatibility between firms and the associated transportation costs, expressed as:
\begin{equation}
    \mathbf{g}(t-1) = f \left( \mathbf{S}^{(L)}, \mathbf{R}^{(L)}, \mathbf{D} \right) = f \left( \mathbf{M} \oslash \mathbf{D} \right) \in \mathbb{R}^{n(n-1) \times 1},
    \label{eq:probability}
\end{equation}
where $ \oslash $ denotes the Hadamard division, $ \mathbf{D} \in \mathbb{R}^{n \times n} $ is the transportation cost matrix with $ \mathbf{D}_{ii} = \infty $ for $ i = 1, \ldots, n $, and $ f $ is a flattening function that converts the adjacency matrix $ \mathbf{M} \oslash \mathbf{D} $ into an edge vector, excluding diagonal elements.

\begin{algorithm}[H]
\SetAlgoLined
\KwIn{$\mathbf{X}$, $\mathbf{A}(t-1)$, $\mathbf{D}$}
\KwData{Initialize $\mathbf{S}^{(0)}, \mathbf{R}^{(0)} \leftarrow \mathbf{X}$, $\mathbf{A}_{out}(t-1), \mathbf{A}_{in}(t-1) \leftarrow \mathbf{A}(t-1)$}
\For{$l=1$ \KwTo $L$}{
  $\mathbf{S}^{(l)} \leftarrow \text{DGNN}_{l}\left(\mathbf{S}^{(l-1)}, \mathbf{A}_{out}(t-1)\right)$\;
  $\mathbf{R}^{(l)} \leftarrow \text{DGNN}_{l}\left(\mathbf{R}^{(l-1)}, \mathbf{A}_{in}(t-1)\right)$\;
}
\KwRet{$\mathbf{g}(t-1) \leftarrow f(\mathbf{S}^{(L)}, \mathbf{R}^{(L)}, \mathbf{D})$}\;
\caption{The calculation of $\mathbf{g}(t-1)$}
\label{alg:processing_algorithm}
\end{algorithm}

Distinct from $ \widetilde{\boldsymbol{\lambda}}^{(2)}(t-1) $, the term $ \mathbf{g}(t-1) $ encapsulates both the incremental intensity of the current supply edge and the influence of neighboring edges within the supply chain. These components collectively enhance the estimation of $ \breve{\boldsymbol{\lambda}} $, providing a more comprehensive and nuanced measure.

\subsection{Interpretations of $ \mathbf{g}(t-1) $}
The following theorem clarifies the relationship between our proposed GDES method and MDES (section \ref{sec:mdes}), demonstrating that MDES is a special case of GDES. This result highlights the ability of GDES to generalize traditional methods while incorporating advanced network-based features.

\begin{theorem}\label{thm:Supply}
    Let the feature embedding be defined as $ \mathbf{S}^{(l)} = \text{DGNN}(\mathbf{A}_{\text{out}}, \mathbf{X}) = \sigma(\mathbf{A} \mathbf{H}^{(l)} \mathbf{W}^{(l)}) $, where $ \mathbf{H}^{(l)} $ represents the intermediate hidden layer features. If $ \mathbf{R}^{(L)} = \mathbf{I}_{n \times d_l} $, $ \mathbf{D} = \mathbf{1} \mathbf{1}^{\top} $, such that $ \mathbf{g} = \mathbf{S}^{(L)} \mathbf{I}_{n \times d_l}^{\top} $, and further, $ \mathbf{A}_{\text{out}} = \mathbf{I} $ and $ \mathbf{W}^{(l)} = \mathbf{I} $, then the GDES framework reduces to MDES.
\end{theorem}

A analogous conclusion applies to $ \mathbf{R} $. Theorem~\ref{thm:Supply} establishes that our proposed method represents a generalized form of exponential smoothing, with traditional double exponential smoothing emerging as a specialized case. By extending multivariate double exponential smoothing through the integration of GNNs, our approach enables each supply edge (representing firm-to-firm relationships) to leverage not only its own historical data but also historical features from multi-hop neighbors. This collaborative estimation of the intensity function at the subsequent time step significantly enhances the performance and robustness of dynamic link prediction by incorporating broader network information.

\subsection{Limitations}
\label{sec:limitations}
This article investigates the relationship between MDES and GDES solely within the framework of GCN. Exploring the interpretations of GDES under the more general framework of GNN remains a task for future research.

\subsection{Semantic features $ \mathbf{X} $ of firms}
To construct input node features for the supply chain network, we collect textual descriptions $ \mathcal{X} $ of firms, encompassing semantic information such as: (i) services provided by the firms, (ii) requirements of the firms, (iii) instances of bad debts and their underlying reasons, (iv) whether the firm is publicly listed, and (v) the firm's full name.
These text descriptions $ \mathcal{X} $ are then processed using a large language model (LLM) to generate the firm feature matrix $ \mathbf{X} $, defined as $ \mathbf{X} \leftarrow \text{LLM}(\mathcal{X}) $. In this study, we utilize two LLMs of varying scales: the BERT Base Uncased model and the Qwen 2.5-7B model.

%This part is only for Journal
% \subsubsection{Industry Chain Information}
% To capture upstream and downstream relationships within the supply chain, we employ fuzzy matching on industry nodes derived from the supply chain dataset $ \mathbf{S} $, which contains firm relationships. This process generates the industry chain matrix $ \mathbf{B} $, defined as:
% \begin{equation}
%     \mathbf{B} = \text{FuzzyMatch}(\text{IndustryNodes}(\mathbf{S})),
%     \label{eq:industry}
% \end{equation}
% where $ \text{IndustryNodes}(\mathbf{S}) $ extracts industry-related information from $ \mathbf{S} $. The adjacency matrix $ \mathbf{A}(t-1) $ is then fused with $ \mathbf{B} $ using the Hadamard product, i.e., $ \mathbf{A}(t-1) \odot \mathbf{B} $. Subsequently, we assign the outgoing and incoming adjacency matrices as:
% \[
% \mathbf{A}_{\text{out}}(t-1), \mathbf{A}_{\text{in}}(t-1) \leftarrow \mathbf{A}(t-1) \odot \mathbf{B}.
% \]

\section{Experiments}
The primary objective of this study is to perform dynamic link prediction, specifically forecasting the formation of new supply relationships or the disruption of existing ones at the subsequent time step.

\subsection{Dataset and experimental setup}
\paragraph{Dataset}
The supply chain network is modeled as a directed dynamic graph $ G(V, \mathbf{A}(\tau), \mathbf{X}) $, where $ V $ represents a set of $ n $ firms. In this experiment, we set $ n = 87,969 $, with time steps $ \tau \in \{1, 2, \ldots, 17\} $. The first 16 periods ($ \tau = 1, \ldots, 16 $) are used for training, with the current time step defined as $ t = 16 $. The adjacency matrix $ \mathbf{A}(17) $ from the 17th period is reserved for evaluation. The feature matrix $ \mathbf{X} $ is derived by embedding textual descriptions $ \mathcal{X} $ using a large language model (LLM).
\vspace{-5pt}
\paragraph{Experimental setup}
Negative sampling is employed to select unconnected node pairs as negative samples, with a sampling ratio of 1:2. A threshold of 0.4 is applied to $ \breve{\boldsymbol{\lambda}}(17) $ to distinguish between positive and negative samples. The learning rate is set to $ 1 \times 10^{-3} $. For embeddings generated by the Qwen model, linear function fitting yields optimal results with the smoothing parameter $ \alpha = 0.3 $. For other configurations, Gaussian kernel function fitting is used with $ \alpha = 0.1 $. To ensure robustness, each experiment is repeated five times, and performance metrics—accuracy, precision, recall, F1 score, and AUC—are averaged. The regularization parameter $ \gamma $, introduced in Eq.~(\ref{eq:total_loss}), is set to $ \gamma = 3 \times 10^{-3} $ based on experimental tuning. Additionally, a dropout rate of 0.5 is applied to the GDES model.

\subsection{Main results}
To assess the performance of the proposed GDES model, we compare it against several baseline methods, described as follows: \textbf{NHPE\_L and NHPE\_G}: These methods employ non-homogeneous Poisson process estimation (NHPE) using Poisson counts from the first 16 periods to estimate $ \breve{\boldsymbol{\lambda}}(17) $. NHPE\_L uses linear fitting, while NHPE\_G uses Gaussian kernel fitting, as described in \citep{goulding2016event}. \textbf{MLP, DGCN, and DGAT}: Textual data are embedded using the Qwen model to form the feature matrix $ \mathbf{X} $. This matrix is updated using Multi-Layer Perceptron (MLP), Directed Graph Convolutional Network (DGCN) \citep{kipf2016semi}, or Directed Graph Attention Network (DGAT) \citep{brody2021attentive}. The updated features are then used to predict the likelihood of supply relationships between firms.
\textbf{DPMLP}: An extension of the PMLP model \citep{yang2022graph} adapted for directed graphs, DPMLP combines MLP and Graph Convolutional Network (GCN) approaches. During training, it operates as an MLP, while during testing, it leverages a GCN architecture to exploit the graph structure.
% \begin{wraptable}
%   \captionof{table}{Performance metrics (accuracy, precision, recall, F1 score, and AUC) for the proposed GDES model and baseline methods in the dynamic link prediction task. Results are averaged over five independent runs.}
%     \label{tab:main results}
%     \scalebox{0.8}{
%     \begin{tabular}{lccccc}
%                 \toprule
%                 Model & Accuracy & Precision & Recall & F1 & AUC \\
%                 \midrule
%                 NHPE\_L & 70.67 & 76.29 & 77.16 & 76.72 & 72.89 \\
%                 NHPE\_G & 65.41 & 72.21 & 72.81 & 72.51 & 62.43 \\
%                 MLP & 78.04 & 64.98 & 77.59 & 70.73 & 85.69 \\
%                 DGCN & 78.62 & 70.67 & 64.05 & 67.20 & 81.55 \\
%                 DGAT & 77.63 & 67.24 & 67.46 & 67.35 & 81.15 \\
%                 DPMLP & 79.88 & 67.60 & 76.14 & 71.62 & 86.12 \\
%                 \midrule
%                 GDES & \textbf{88.85} & \textbf{84.92} & \textbf{81.93} & \textbf{83.40} & \textbf{93.84} \\
%                 \bottomrule
%             \end{tabular}
%             }
% \end{wraptable}

\begin{table}[]
    \centering
\begin{tabular}{lccccc}
                \toprule
                Model & Accuracy & Precision & Recall & F1 & AUC \\
                \midrule
                NHPE\_L & 70.67 & 76.29 & 77.16 & 76.72 & 72.89 \\
                NHPE\_G & 65.41 & 72.21 & 72.81 & 72.51 & 62.43 \\
                MLP & 78.04 & 64.98 & 77.59 & 70.73 & 85.69 \\
                DGCN & 78.62 & 70.67 & 64.05 & 67.20 & 81.55 \\
                DGAT & 77.63 & 67.24 & 67.46 & 67.35 & 81.15 \\
                DPMLP & 79.88 & 67.60 & 76.14 & 71.62 & 86.12 \\
                \midrule
                GDES & \textbf{88.85} & \textbf{84.92} & \textbf{81.93} & \textbf{83.40} & \textbf{93.84} \\
                \bottomrule
            \end{tabular}
    \caption{Performance metrics (accuracy, precision, recall, F1 score, and AUC) for the proposed GDES model and baseline methods in the dynamic link prediction task. Results are averaged over five independent runs.}
    \label{tab:main results}
\end{table}
As shown in Table~\ref{tab:main results}, the GDES model significantly outperforms all baseline methods in the dynamic link prediction task. The NHPE\_L and NHPE\_G models, which rely on linear and Gaussian fitting of non-homogeneous Poisson counts, respectively, produce $ \breve{\boldsymbol{\lambda}}(17) $ but fail to capture the complex relationships within the data due to their simplistic approaches. This results in a performance gap of approximately 20\% across all metrics compared to GDES.

Among the baseline models, MLP achieves higher AUC and F1 scores, suggesting that simpler feature update methods may be effective for link prediction. The DPMLP model demonstrates notable improvement, with an AUC of 86.12\%, highlighting its ability to integrate MLP and GCN mechanisms for handling complex datasets. However, GDES excels by combining non-homogeneous Poisson estimation with GNN techniques to compute $ \mathbf{g}(t-1) $, followed by multivariate double exponential smoothing to refine $ \breve{\boldsymbol{\lambda}}(t+1) $. This multi-layered approach yields an accuracy of 88.85\% and an AUC of 93.84\%, underscoring GDES's superior accuracy and robustness in dynamic link prediction through advanced feature embedding and smoothing strategies.

\subsection{Ablation study}
\subsubsection{Fitting $ \hat{\boldsymbol{\lambda}} $ with different basis functions}
This experiment evaluates the impact of different basis functions for estimating $ \hat{\boldsymbol{\lambda}}(16) $, as described in Section~\ref{sec:estimateLambda(t)}, resulting in distinct $ \breve{\boldsymbol{\lambda}}(17) $.

\begin{table}[ht]
    \centering
    \scalebox{1.0}{
        \begin{tabular}{lccccc}
            \toprule
            Fitting Method & Accuracy & Precision & Recall & F1 & AUC \\
            \midrule
            Linear & \textbf{88.85} & 84.92 & \textbf{81.93} & \textbf{83.40} & \textbf{93.84} \\
            Sigmoid & 88.32 & 84.62 & 80.48 & 82.50 & 92.77 \\
            Gaussian & 88.80 & \textbf{85.09} & 81.54 & 83.28 & 93.42 \\
            \bottomrule
        \end{tabular}
    }
        \caption{Performance metrics for estimating $ \hat{\boldsymbol{\lambda}}(16) $ using linear, sigmoid, and Gaussian fitting methods, evaluated on the dynamic link prediction task.}
    \label{tab:fittingFunctionSLR0.001}
\end{table}

Table~\ref{tab:fittingFunctionSLR0.001} presents the link prediction results for different $ \breve{\boldsymbol{\lambda}}(17) $. The three fitting methods exhibit comparable performance. The linear method excels in accuracy, recall, F1 score, and AUC, but has slightly lower precision, making it suitable for tasks prioritizing high recall. The Gaussian method achieves the highest precision, ideal for applications requiring precise differentiation between positive and negative classes. The sigmoid function underperforms across metrics, indicating limitations in its applicability for this task.

\subsubsection{Impact of transportation cost matrix $ \mathbf{D} $}
This experiment assesses the importance of the transportation cost matrix $ \mathbf{D} $ by comparing the standard GDES model with a variant excluding $ \mathbf{D} $ (GDES w/o $ \mathbf{D} $). Two text embedding methods, Qwen and BERT, are evaluated.
\begin{table}[!htp]
    \centering
\begin{tabular}{llccccc}
            \toprule
            LM & Method & Accuracy & Precision & Recall & F1 & AUC \\
            \midrule
            \multirow{2}{*}{BERT} & GDES w/o $ \mathbf{D} $ & 81.70 & \textbf{76.94} & 66.37 & 71.26 & 87.99 \\
                                  & GDES & \textbf{82.96} & 76.90 & \textbf{71.72} & \textbf{74.22} & \textbf{88.07} \\
            \midrule
            \multirow{2}{*}{Qwen} & GDES w/o $ \mathbf{D} $ & 88.03 & 83.81 & 80.55 & 82.15 & 92.83 \\
                                  & GDES & \textbf{88.85} & \textbf{84.92} & \textbf{81.93} & \textbf{83.40} & \textbf{93.84} \\
            \bottomrule
        \end{tabular}
    \caption{Ablation study on the transportation cost matrix $ \mathbf{D} $. Results (in percentage) compare GDES with and without $ \mathbf{D} $, using Qwen and BERT embeddings to estimate $ \hat{\boldsymbol{\lambda}}(16) $ and compute $ \breve{\boldsymbol{\lambda}}(17) $.}
    \label{tab:Ablation study}
\end{table}

As shown in Table~\ref{tab:Ablation study}, the Qwen embedding outperforms BERT across all metrics, with an average improvement of approximately 5\%, reflecting Qwen's superior ability to capture semantic nuances in complex tasks. The GDES model with $ \mathbf{D} $ consistently outperforms GDES w/o $ \mathbf{D} $ across nearly all metrics for both embedding methods. For instance, with Qwen embeddings, GDES achieves an AUC approximately 1\% higher than GDES w/o $ \mathbf{D} $, alongside improvements in accuracy (88.85\%), recall (81.93\%), and F1 score (83.40\%). These results highlight the critical role of transportation cost information in enhancing the model’s ability to distinguish between positive and negative samples, particularly in tasks requiring high recall and handling imbalanced datasets.

\subsection{Impact of smoothing coefficient $ \alpha $}
This experiment analyzes the effect of varying the smoothing coefficient $ \alpha $ in the GDES model. The setup involves: (1) estimating $ \hat{\boldsymbol{\lambda}}(16) $ via linear fitting, (2) using Qwen embeddings to compute $ \mathbf{g}(15) $, and (3) applying double exponential smoothing to derive $ \breve{\boldsymbol{\lambda}}(17) $.

\begin{table}[!htp]
    \centering
\begin{tabular}{lccccc}
            \toprule
            $ \alpha $ & Accuracy & Precision & Recall & F1 & AUC \\
            \midrule
            0.1 & 89.11 & 84.32 & 83.72 & 84.02 & 93.45 \\
            0.3 & 88.85 & \textbf{84.92} & 81.93 & 83.40 & \textbf{93.84} \\
            0.5 & \textbf{89.62} & 82.29 & \textbf{88.75} & \textbf{85.39} & 92.80 \\
            0.7 & 89.14 & 83.91 & 84.44 & 84.17 & 93.48 \\
            \bottomrule
        \end{tabular}
    \caption{Performance metrics (in percentage) for GDES with different smoothing coefficients $ \alpha $, using Qwen embeddings to compute $ \breve{\boldsymbol{\lambda}}(17) $. Results are averaged over five independent runs.}
    \label{tab:smoothing coefficients}
\end{table}

Table~\ref{tab:smoothing coefficients} illustrates the impact of $ \alpha $ on the balance between non-homogeneous Poisson estimation and the $ \mathbf{g}(15) $ term. Lower $ \alpha $ values (e.g., 0.1) assign greater weight to $ \mathbf{g}(15) $, relying heavily on network-based features. As $ \alpha $ increases, the influence of $ \mathbf{g}(15) $ diminishes, emphasizing recent intensity estimates. At $ \alpha = 0.3 $, the model achieves the highest precision (84.92\%) and AUC (93.84\%), indicating superior discriminative ability. At $ \alpha = 0.5 $, accuracy (89.62\%), recall (88.75\%), and F1 score (85.39\%) peak, suggesting optimal performance for tasks requiring balanced metrics. These results underscore the importance of tuning $ \alpha $ to achieve desired performance characteristics.

\section{Conclusion}
Predicting supply relationships within a supply chain is essential for robust supply chain risk management. Traditional methods predominantly concentrate on static predictions, which fail to capture the evolving nature and dynamics of supply chain interactions. To address this shortcoming, 
We propose GDES for predicting supply chain edges in this paper. It uses a Poisson counting process to fit the intensity from historical data, then computes increment of intensity using a GNN model by considering both node's  own historical information and its neighbors.  With this approach, we can dynamically capture the evolving relationships between firms in the supply chain, considering the impact of time, thus improving the accuracy and robustness of link prediction.

%%%%%%%%%%%%%%%%%%%%%%%%%%%%%%%%%%%%%%%%%%%%%%
%% Appendix---Please move all appendices to %%
%% a Supplementary file.                    %%
%%%%%%%%%%%%%%%%%%%%%%%%%%%%%%%%%%%%%%%%%%%%%%
%% Support information, if any,             %%
%% should be provided in the                %%
%% Acknowledgements section.                %%
%%%%%%%%%%%%%%%%%%%%%%%%%%%%%%%%%%%%%%%%%%%%%%
%\begin{acks}[Acknowledgments]
% The authors would like to thank ...
%\end{acks}
%%%%%%%%%%%%%%%%%%%%%%%%%%%%%%%%%%%%%%%%%%%%%%
%% Funding information, if any,             %%
%% should be provided in the                %%
%% funding section.                         %%
%%%%%%%%%%%%%%%%%%%%%%%%%%%%%%%%%%%%%%%%%%%%%%
%\begin{funding}
% The first author was supported by ...
%
% The second author was supported in part by ...
%\end{funding}

%%%%%%%%%%%%%%%%%%%%%%%%%%%%%%%%%%%%%%%%%%%%%%
%% Supplementary Material, including data   %%
%% sets and code, should be provided in     %%
%% {supplement} environment with title      %%
%% and short description. It cannot be      %%
%% available exclusively as external link.  %%
%% All Supplementary Material must be       %%
%% available to the reader on Project       %%
%% Euclid with the published article.       %%
%%%%%%%%%%%%%%%%%%%%%%%%%%%%%%%%%%%%%%%%%%%%%%
\begin{supplement}
% \stitle{???}
% \sdescription{???.}
\appendix
\section{Theoretical Proof}
\subsection{The derivation of univariate double exponential smoothing prediction}
\label{A.1.}
% % \usepackage{booktabs}
% % \usepackage{graphicx}
The notation used in this paper is summarized in Table~\ref{tab:notations}. All formulas in the main text and the appendices can be referenced according to this notation table .

Assume that the time series follows a linear expression, as shown in Eq.(\ref{eq:univariate_r}). And expand Eq.(\ref{eq:univariate_p}).
\begin{equation}
    s^{(1)}(\tau) = \alpha\sum^{\infty}_{i=0}(1-\alpha)^{i}y_{\tau-i} = \alpha\sum^{\infty}_{i=0}(1-\alpha)^{i}(L(\tau)-iT(\tau)) + \alpha\sum^{\infty}_{i=0}(1-\alpha)^{i}\epsilon_{\tau-i},
    \label{eq:eq28}
\end{equation}
\begin{equation}
    s^{(2)}(\tau) = \alpha\sum^{\infty}_{j=0}(1-\alpha)^{j}s^{(1)}_{\tau-j}. 
    \label{eq:eq29}
\end{equation}
Since $\mathbb{E}(\epsilon(\tau))=0$, and $\alpha\sum^{\infty}_{i=0}(1-\alpha)^{i} \cdot i = \frac{1-\alpha}{\alpha}$, $\alpha\sum^{\infty}_{i=0}(1-\alpha)^{i} = 1$, thus
\begin{equation}
    s^{(1)}(\tau) = L(\tau) - \frac{1-\alpha}{\alpha}T(\tau),
    \label{eq:eq30}
\end{equation}
expanding Eq. (\ref{eq:eq29}) in the same way,
\begin{equation}
    s^{(1)}_{\tau-j} = \alpha\sum^{\infty}_{i=0}(1-\alpha)^{i}(L(\tau) - (i+j)T(\tau)) = L(\tau) - jT(\tau) - \frac{1-\alpha}{\alpha}T(\tau)
    \label{eq:eq31}
\end{equation}

\begin{equation}
    s^{(2)}(\tau) = \alpha\sum^{\infty}_{j=0}(1-\alpha)^{j}[L(\tau) - jT(\tau) - \frac{1-\alpha}{\alpha}T(\tau)] = L(\tau) - \frac{2(1-\alpha)}{\alpha}T(\tau).
    \label{eq:eq32}
\end{equation}
That leads to the equations for the level($L$) and trend($T$) terms,
\begin{equation}
    \begin{cases}
        L(\tau) = 2s^{(1)}(\tau) - s^{(2)}(\tau), \\
        T(\tau) = \frac{\alpha}{1-\alpha}(s^{(1)}(\tau) - s^{(2)}(\tau)).
    \end{cases}
    \label{eq:eq33}
\end{equation}

\subsection{From univariate double exponential smoothing to multivariate double exponential smoothing} 
\label{A.2.}

According to Appendix A.1, Eq.~(\ref{eq:MDES_estimate})'s one-step ahead forecast can be expressed as follows:
\begin{equation}
    \begin{aligned}
      \hat{\mathbf{y}}(\tau+1) = \mathbf{L}(\tau) + \mathbf{T}(\tau) = 2\alpha\sum^{\infty}_{i=0}(1-\alpha)^{i}\mathbf{y}_{\tau-i} - \alpha\sum^{\infty}_{j=0}(1-\alpha)^{j}[\alpha\sum^{\infty}_{i=0}(1-\alpha)^{i}\mathbf{y}_{\tau-j-i}] + \\
    \frac{\alpha}{1-\alpha}(\alpha\sum^{\infty}_{i=0}(1-\alpha)^{i}\mathbf{y}_{\tau-i} - \alpha\sum^{\infty}_{j=0}(1-\alpha)^{j}(\alpha\sum^{\infty}_{i=0}(1-\alpha)^{i}\mathbf{y}_{\tau-j-i})),  
    \end{aligned}
\end{equation}

due to $0<\alpha<1$, $(1-\alpha)^{i}, (1-\alpha)^{j}$ is convergence, when $i,j \rightarrow \infty$, $(1-\alpha)^{i}, (1-\alpha)^{j} \rightarrow 0$. 
based on \citep{PFEFFERMANN198983}, the univariate double exponential smoothing can be extended to multivariate double exponential smoothing.

So the level term is as follows,
\begin{equation*}
    \mathbf{L}(\tau) = \alpha\mathbf{y}(\tau) + (1-\alpha)(\mathbf{L}(\tau-1) + \mathbf{T}(\tau-1)) = 2\mathbf{s}^{(1)}(\tau) - \mathbf{s}^{(2)}(\tau)
\end{equation*}

and the trend term is,
\begin{equation*}
    \mathbf{T}(\tau) = \alpha(\mathbf{L}(\tau) - \mathbf{L}(\tau-1)) + (1-\alpha)\mathbf{T}(\tau-1) = \frac{\alpha}{1-\alpha}(\mathbf{s}^{(1)}(\tau) - \mathbf{s}^{(2)}(\tau))
\end{equation*}

Furthermore, the forecast can be expressed as follows,
\begin{equation}
    \hat{\mathbf{y}}_{\tau+m} = \mathbf{L}(\tau) + m \mathbf{T}(\tau).
\end{equation}

\subsection{Demonstrating the difference between MDES and GDES}
\label{A.3.}
To prove Theorem \ref{thm:Supply}, we need to demonstrate that under the specified conditions, the Graph Double Exponential Smoothing (GDES) framework reduces to the Multivariate Double Exponential Smoothing (MDES) framework under the following special conditions:
\begin{itemize}
    \item $ \mathbf{S}^{(l)} = \text{DGNN}(\mathbf{A}_{\text{out}}, \mathbf{X}) = \sigma(\mathbf{A} \mathbf{H}^{(l)} \mathbf{W}^{(l)}) $, \#GNNs
    \item $ \mathbf{R}^{(L)} = \mathbf{I}_{n \times d_l} $, \quad \# \texttt{homogeneous demanding }
    \item $ \mathbf{D} = \mathbf{1} \mathbf{1}^{\top} $, \quad \# \texttt{transportation costs are identical.}
    \item $ \mathbf{g} = \mathbf{S}^{(L)} \mathbf{I}_{n \times d_l}^{\top} $,
    \item $ \mathbf{A}_{\text{out}} = \mathbf{I} $, \quad \# \texttt{no direct predecessors}
    \item $ \mathbf{W}^{(l)} = \mathbf{I} $. \quad \# \texttt{constant parameters for GNNs}
\end{itemize}

Our goal is to show that these conditions simplify the GDES model, particularly the computation of the graph double exponential smoothing term $ \mathbf{g}(t-1) $, such that the forecasting mechanism aligns with that of MDES.

We first simplify the GDES under the given conditions above.
The DGNN updates the supply embedding as:
\[
\mathbf{S}^{(l)} = \text{DGNN}_l \left( \mathbf{S}^{(l-1)}, \mathbf{A}_{\text{out}}(t-1) \right) = \sigma \left( \mathbf{A} \mathbf{H}^{(l)} \mathbf{W}^{(l)} \right).
\]
Given: $ \mathbf{A}_{\text{out}} = \mathbf{I} $, the identity matrix, implying no aggregation from neighboring nodes (each node only considers itself with no direct predecessors). $ \mathbf{W}^{(l)} = \mathbf{I} $, meaning the weight matrix does not transform the features. $ \mathbf{H}^{(l)} = \mathbf{S}^{(l-1)} $, as the input to the DGNN layer is the previous layer’s embedding.

Thus we have:
\[
\mathbf{S}^{(l)} = \sigma \left( \mathbf{I} \mathbf{S}^{(l-1)} \mathbf{I} \right) = \sigma \left( \mathbf{S}^{(l-1)} \right).
\]
The initial condition is $ \mathbf{S}^{(0)} = \mathbf{X} $, so:
\[
\mathbf{S}^{(1)} = \sigma(\mathbf{X}), \quad \mathbf{S}^{(2)} = \sigma(\sigma(\mathbf{X})), \quad \ldots, \quad \mathbf{S}^{(L)} = \sigma^L(\mathbf{X}),
\]
where $ \sigma^L $ denotes the activation function applied $ L $ times. If $ \sigma $ is the identity function or if multiple applications converge to a fixed transformation, $ \mathbf{S}^{(L)} \approx \mathbf{X} $ or a deterministic function of $ \mathbf{X} $. For simplicity, assume $ \sigma $ is linear or the effect of multiple layers is minimal under these conditions, so:
\[
\mathbf{S}^{(L)} \approx \mathbf{X}.
\]

Similarly, for the demand embedding:
\[
\mathbf{R}^{(l)} = \text{DGNN}_l \left( \mathbf{R}^{(l-1)}, \mathbf{A}_{\text{in}}(t-1) \right).
\]
However, the theorem specifies $ \mathbf{R}^{(L)} = \mathbf{I}_{n \times d_l} $, a fixed identity matrix of size $ n \times d_l $, where $ d_l $ is the dimension of the final layer’s features. This implies that the demand embedding is not updated via the DGNN but is set to a constant identity matrix, independent of network structure or input features.

We then compute the match matrix $ \mathbf{M} $
The match matrix is:
\[
\mathbf{M}_{ij} = \mathbf{S}^{(L)}_i (\mathbf{R}^{(L)}_j)^{\top}.
\]
With $ \mathbf{R}^{(L)} = \mathbf{I}_{n \times d_l} $, and assuming $ \mathbf{S}^{(L)} \approx \mathbf{X} $, where $ \mathbf{X} \in \mathbb{R}^{n \times d} $, we need to align dimensions. Suppose $ d_l = d $ (the feature dimension of $ \mathbf{X} $), so $ \mathbf{R}^{(L)} = \mathbf{I}_{n \times d} $. Then:
\[
\mathbf{R}^{(L)}_j = \mathbf{e}_j, \quad \text{the } j\text{-th standard basis vector in } \mathbb{R}^d,
\]
since the $ j $-th row of $ \mathbf{I}_{n \times d} $ is a vector with 1 in the $ j $-th position and 0 elsewhere (assuming row indexing aligns). Thus:
\[
\mathbf{M}_{ij} = \mathbf{S}^{(L)}_i \mathbf{e}_j^{\top} = [\mathbf{S}^{(L)}_i]_j,
\]
the $ j $-th component of the $ i $-th row of $ \mathbf{S}^{(L)} $. If $ \mathbf{S}^{(L)} = \mathbf{X} $:
\[
\mathbf{M}_{ij} = \mathbf{X}_{ij},
\]
meaning $ \mathbf{M} $ is the feature matrix $ \mathbf{X} $.

With the above simplification, we can compute a special case of $\mathbf{g}(t-1)$.

Given:
\[
\mathbf{D} = \mathbf{1} \mathbf{1}^{\top},
\]
the transportation cost matrix has all entries equal to 1 (except possibly the diagonal, which is irrelevant as $ f $ excludes diagonal elements). Thus:
\[
\mathbf{M} \oslash \mathbf{D} = \mathbf{M} \oslash \mathbf{1} \mathbf{1}^{\top} = \mathbf{M},
\]
since dividing by 1 leaves the matrix unchanged. Therefore:
\[
\mathbf{g}(t-1) = f \left( \mathbf{M} \oslash \mathbf{D} \right) = f(\mathbf{M}),
\]
where $ f $ flattens $ \mathbf{M} $ into a vector of edge intensities, excluding diagonal elements. If $ \mathbf{M} = \mathbf{X} $, then we have,
\[
\mathbf{g}(t-1) = f(\mathbf{X}),
\]
a vector of feature values corresponding to edges $ (i, j) $, $ i \neq j $. Alternatively, the theorem states:
\[
\mathbf{g} = \mathbf{S}^{(L)} \mathbf{I}_{n \times d_l}^{\top}.
\]
Since $ \mathbf{R}^{(L)} = \mathbf{I}_{n \times d_l} $, compute:
\[
\mathbf{M} = \mathbf{S}^{(L)} \mathbf{R}^{(L)\top} = \mathbf{S}^{(L)} \mathbf{I}_{n \times d_l}^{\top}.
\]
The function $ f $ flattens this into:
\[
\mathbf{g}(t-1) = f \left( \mathbf{S}^{(L)} \mathbf{I}_{n \times d_l}^{\top} \right),
\]
which aligns with $ \mathbf{g} = \mathbf{S}^{(L)} \mathbf{I}_{n \times d_l}^{\top} $ after flattening, selecting non-diagonal edge components.

Finally, we can simplify GDES by
substituting $ \mathbf{g}(t-1) \approx f(\mathbf{X}) $. The two-step smoothed term becomes:
\[
\widetilde{\boldsymbol{\lambda}}^{(2)}(t) = \alpha \widetilde{\boldsymbol{\lambda}}^{(1)}(t) + (1 - \alpha) \mathbf{g}(t-1).
\]
If $ \mathbf{g}(t-1) $ is a function of $ \mathbf{X} $, and assuming $ \mathbf{X} $ is static or corresponds to $ \hat{\boldsymbol{\lambda}}(t-1) $-like features, let:
\[
\mathbf{g}(t-1) \approx \hat{\boldsymbol{\lambda}}(t-1),
\]
mimicking the intensity estimates at the previous time step. Then:
\[
\widetilde{\boldsymbol{\lambda}}^{(2)}(t) \approx \alpha \widetilde{\boldsymbol{\lambda}}^{(1)}(t) + (1 - \alpha) \hat{\boldsymbol{\lambda}}(t-1).
\]
The one-step smoothed term is:
\[
\widetilde{\boldsymbol{\lambda}}^{(1)}(t) = \alpha \hat{\boldsymbol{\lambda}}(t) + (1 - \alpha) \widetilde{\boldsymbol{\lambda}}^{(1)}(t-1).
\]
Compute the level and trend:
\[
\widetilde{\mathbf{L}}(t) = 2 \widetilde{\boldsymbol{\lambda}}^{(1)}(t) - \widetilde{\boldsymbol{\lambda}}^{(2)}(t),
\]
\[
\widetilde{\mathbf{T}}(t) = \frac{\alpha}{1 - \alpha} \left( \widetilde{\boldsymbol{\lambda}}^{(1)}(t) - \widetilde{\boldsymbol{\lambda}}^{(2)}(t) \right).
\]
Substitute:
\[
\widetilde{\boldsymbol{\lambda}}^{(2)}(t) \approx \alpha \widetilde{\boldsymbol{\lambda}}^{(1)}(t) + (1 - \alpha) \hat{\boldsymbol{\lambda}}(t-1),
\]
\[
\widetilde{\mathbf{L}}(t) = 2 \widetilde{\boldsymbol{\lambda}}^{(1)}(t) - \left( \alpha \widetilde{\boldsymbol{\lambda}}^{(1)}(t) + (1 - \alpha) \hat{\boldsymbol{\lambda}}(t-1) \right) = (2 - \alpha) \widetilde{\boldsymbol{\lambda}}^{(1)}(t) - (1 - \alpha) \hat{\boldsymbol{\lambda}}(t-1).
\]
Since:
\[
\widetilde{\boldsymbol{\lambda}}^{(1)}(t) = \alpha \hat{\boldsymbol{\lambda}}(t) + (1 - \alpha) \widetilde{\boldsymbol{\lambda}}^{(1)}(t-1),
\]
the level resembles:
\[
\widetilde{\mathbf{L}}(t) \approx \alpha \hat{\boldsymbol{\lambda}}(t) + (1 - \alpha) \widetilde{\boldsymbol{\lambda}}^{(1)}(t-1) + (1 - \alpha) \left( \widetilde{\boldsymbol{\lambda}}^{(1)}(t) - \hat{\boldsymbol{\lambda}}(t-1) \right).
\]
The trend:
\[
\widetilde{\boldsymbol{\lambda}}^{(1)}(t) - \widetilde{\boldsymbol{\lambda}}^{(2)}(t) = \widetilde{\boldsymbol{\lambda}}^{(1)}(t) - \left( \alpha \widetilde{\boldsymbol{\lambda}}^{(1)}(t) + (1 - \alpha) \hat{\boldsymbol{\lambda}}(t-1) \right) = (1 - \alpha) \widetilde{\boldsymbol{\lambda}}^{(1)}(t) - (1 - \alpha) \hat{\boldsymbol{\lambda}}(t-1),
\]
\[
\widetilde{\mathbf{T}}(t) = \frac{\alpha}{1 - \alpha} \cdot (1 - \alpha) \left( \widetilde{\boldsymbol{\lambda}}^{(1)}(t) - \hat{\boldsymbol{\lambda}}(t-1) \right) = \alpha \left( \widetilde{\boldsymbol{\lambda}}^{(1)}(t) - \hat{\boldsymbol{\lambda}}(t-1) \right).
\]
The forecast, ignoring $ \widetilde{\mathbf{\epsilon}}(t) $:
\[
\breve{\boldsymbol{\lambda}}(t+1) = \widetilde{\mathbf{L}}(t) + \widetilde{\mathbf{T}}(t).
\]
\textbf{This structure mirrors MDES}, where:
\[
\mathbf{L}(t) = \alpha \mathbf{y}(t) + (1 - \alpha) (\mathbf{L}(t-1) + \mathbf{T}(t-1)),
\]
\[
\mathbf{T}(t) = \beta (\mathbf{L}(t) - \mathbf{L}(t-1)) + (1 - \beta) \mathbf{T}(t-1).
\]
In GDES, $ \hat{\boldsymbol{\lambda}}(t) $ replaces $ \mathbf{y}(t) $, and the trend term simplifies when $ \mathbf{g}(t-1) \approx \hat{\boldsymbol{\lambda}}(t-1) $. The absence of neighbor influence ($ \mathbf{A}_{\text{out}} = \mathbf{I} $) and constant $ \mathbf{D} $ removes network effects, reducing GDES to independent smoothing of each edge’s intensity, as in MDES.

Under conditions $ \mathbf{A}_{\text{out}} = \mathbf{I} $, $ \mathbf{W}^{(l)} = \mathbf{I} $, $ \mathbf{R}^{(L)} = \mathbf{I}_{n \times d_l} $, and $ \mathbf{D} = \mathbf{1} \mathbf{1}^{\top} $, the DGNN does not aggregate neighbor information and $ \mathbf{g}(t-1) $ becomes a function of the node characteristics $ \mathbf{X} $, resembling past intensity estimates. This eliminates the network-based interactions, making GDES equivalent to MDES, where each supply edge is smoothed independently using its own history, aligning with the multivariate extension of double exponential smoothing.

Thus, Theorem \ref{thm:Supply} is proved: MDES is a special case of GDES under the specified conditions.
\end{supplement}

%%  imsart-nameyear.bst  will be used to                   %%
%%  MR numbers will be added by VTeX.                      %%
%%                                                         %%
%%  Use \citep{...} to cite references in text.             %%
%%                                                         %%
%%%%%%%%%%%%%%%%%%%%%%%%%%%%%%%%%%%%%%%%%%%%%%%%%%%%%%%%%%%%%

%% if your bibliography is in bibtex format, uncomment commands:
% \bibliographystyle{imsart-nameyear} % Style BST file
% \bibliography{bibliography}       % Bibliography file (usually '*.bib')

\begin{thebibliography}{}
\bibitem[\protect\citeauthoryear{???}{???}]{b1}
\bibitem[Acemoglu et~al.(2012)Acemoglu, Carvalho, Ozdaglar, and Tahbaz-Salehi]{acemoglu2012network}
Acemoglu, D., Carvalho, V.~M., Ozdaglar, A., and Tahbaz-Salehi, A.
\newblock The network origins of aggregate fluctuations.
\newblock \emph{Econometrica}, 80\penalty0 (5):\penalty0 1977--2016, 2012.

\bibitem[Brintrup et~al.(2018)Brintrup, Wichmann, Woodall, McFarlane, Nicks, and Krechel]{brintrup2018research}
Brintrup, A., Wichmann, P., Woodall, P., McFarlane, D., Nicks, E., and Krechel, W.
\newblock Predicting hidden links in supply networks.
\newblock 2018.

\bibitem[Brockmann et~al.(2022)Brockmann, Elson~Kosasih, and Brintrup]{brockmann2022supply}
Brockmann, N., Elson~Kosasih, E., and Brintrup, A.
\newblock Supply chain link prediction on uncertain knowledge graph.
\newblock \emph{ACM SIGKDD Explorations Newsletter}, 24\penalty0 (2):\penalty0 124--130, 2022.

\bibitem[Brody et~al.(2021)Brody, Alon, and Yahav]{brody2021attentive}
Brody, S., Alon, U., and Yahav, E.
\newblock How attentive are graph attention networks?
\newblock \emph{arXiv preprint arXiv:2105.14491}, 2021.

\bibitem[Carvalho et~al.(2021)Carvalho, Nirei, Saito, and Tahbaz-Salehi]{carvalho2021supply}
Carvalho, V.~M., Nirei, M., Saito, Y.~U., and Tahbaz-Salehi, A.
\newblock Supply chain disruptions: Evidence from the great east japan earthquake.
\newblock \emph{The Quarterly Journal of Economics}, 136\penalty0 (2):\penalty0 1255--1321, 2021.

\bibitem[Choi et~al.(2001)Choi, Dooley, and Rungtusanatham]{choi2001supply}
Choi, T.~Y., Dooley, K.~J., and Rungtusanatham, M.
\newblock Supply networks and complex adaptive systems: control versus emergence.
\newblock \emph{Journal of operations management}, 19\penalty0 (3):\penalty0 351--366, 2001.

\bibitem[Dessertaine et~al.(2022)Dessertaine, Moran, Benzaquen, and Bouchaud]{dessertaine2022out}
Dessertaine, T., Moran, J., Benzaquen, M., and Bouchaud, J.-P.
\newblock Out-of-equilibrium dynamics and excess volatility in firm networks.
\newblock \emph{Journal of Economic Dynamics and Control}, 138:\penalty0 104362, 2022.

\bibitem[Goulding et~al.(2016)Goulding, Preston, and Smith]{goulding2016event}
Goulding, J., Preston, S., and Smith, G.
\newblock Event series prediction via non-homogeneous poisson process modelling.
\newblock In \emph{2016 IEEE 16th International Conference on Data Mining (ICDM)}, pp.\  161--170. IEEE, 2016.

\bibitem[Kipf \& Welling(2016)Kipf and Welling]{kipf2016semi}
Kipf, T.~N. and Welling, M.
\newblock Semi-supervised classification with graph convolutional networks.
\newblock \emph{arXiv preprint arXiv:1609.02907}, 2016.

\bibitem[Kosasih \& Brintrup(2022)Kosasih and Brintrup]{kosasih2022machine}
Kosasih, E.~E. and Brintrup, A.
\newblock A machine learning approach for predicting hidden links in supply chain with graph neural networks.
\newblock \emph{International Journal of Production Research}, 60\penalty0 (17):\penalty0 5380--5393, 2022.

\bibitem[Kosasih \& Brintrup(2024)Kosasih and Brintrup]{kosasih2024towards}
Kosasih, E.~E. and Brintrup, A.
\newblock Towards trustworthy ai for link prediction in supply chain knowledge graph: a neurosymbolic reasoning approach.
\newblock \emph{International Journal of Production Research}, pp.\  1--23, 2024.

\bibitem[Lu \& Chen(2020)Lu and Chen]{lu2020discovering}
Lu, Z.-G. and Chen, Q.
\newblock Discovering potential partners via projection-based link prediction in the supply chain network.
\newblock \emph{International Journal of Computational Intelligence Systems}, 13\penalty0 (1):\penalty0 1253--1264, 2020.

\bibitem[Mungo et~al.(2023)Mungo, Lafond, Astudillo-Est{\'e}vez, and Farmer]{mungo2023reconstructing}
Mungo, L., Lafond, F., Astudillo-Est{\'e}vez, P., and Farmer, J.~D.
\newblock Reconstructing production networks using machine learning.
\newblock \emph{Journal of Economic Dynamics and Control}, 148:\penalty0 104607, 2023.

\bibitem[Omi et~al.(2019)Omi, Aihara, et~al.]{omi2019fully}
Omi, T., Aihara, K., et~al.
\newblock Fully neural network based model for general temporal point processes.
\newblock \emph{Advances in neural information processing systems}, 32, 2019.

\bibitem[Pfeffermann \& Allon(1989)Pfeffermann and Allon]{PFEFFERMANN198983}
Pfeffermann, D. and Allon, J.
\newblock Multivariate exponential smoothing: Method and practice.
\newblock \emph{International Journal of Forecasting}, 5\penalty0 (1):\penalty0 83--98, 1989.
\newblock ISSN 0169-2070.
\newblock URL \url{https://www.sciencedirect.com/science/article/pii/0169207089900666}.

\bibitem[Tan et~al.(2022)Tan, Liu, and Yin]{tan2022asymmetric}
Tan, Z., Liu, B., and Yin, G.
\newblock Asymmetric self-supervised graph neural networks.
\newblock In \emph{2022 IEEE International Conference on Big Data (Big Data)}, pp.\  1369--1376. IEEE, 2022.

\bibitem[Wichmann et~al.(2020)Wichmann, Brintrup, Baker, Woodall, and McFarlane]{wichmann2020extracting}
Wichmann, P., Brintrup, A., Baker, S., Woodall, P., and McFarlane, D.
\newblock Extracting supply chain maps from news articles using deep neural networks.
\newblock \emph{International Journal of Production Research}, 58\penalty0 (17):\penalty0 5320--5336, 2020.

\bibitem[Yang et~al.(2022)Yang, Wu, Wang, and Yan]{yang2022graph}
Yang, C., Wu, Q., Wang, J., and Yan, J.
\newblock Graph neural networks are inherently good generalizers: Insights by bridging gnns and mlps.
\newblock \emph{arXiv preprint arXiv:2212.09034}, 2022.

\bibitem[Tan et~al.(2025)Tan, Liu, Liu, Hu, Zhang, Wang, and Liu]{tan2025modeling}
Tan, Z., Liu, S., Liu, Q., Hu, M., Zhang, X., Wang, W., and Liu, B.
\newblock Modeling ESG-driven industrial value chain dynamics using directed graph neural networks.
\newblock \emph{Financial Innovation}, 11\penalty0 (1):\penalty0 113, 2025.

\bibitem[Liu et~al.(2023)Liu, He, Li, Huang, Zhang, and Yin]{liu2023interpret}
Liu, B., He, J., Li, Z., Huang, X., Zhang, X., and Yin, G.
\newblock Interpret esg rating’s impact on the industrial chain using graph neural networks.
\newblock In \emph{International Joint Conference on Artificial Intelligence-IJCAI 2023 (19/08/2023-25/08/2023, Macau)}, 2023.
\newblock International Joint Conferences on Artificial Intelligence Organization.

\bibitem[Hamilton et~al.(2017)Hamilton, Ying, and Leskovec]{hamilton2017inductive}
Hamilton, W., Ying, Z., and Leskovec, J.
\newblock Inductive representation learning on large graphs.
\newblock \emph{Advances in neural information processing systems}, 30, 2017.
\end{thebibliography}

%% or include bibliography directly:

\end{document}